\documentclass[11pt]{article}
\usepackage{epstopdf}
\usepackage[a4paper,hmarginratio=1:1,vmarginratio=2:3,totalwidth=15.3cm,totalheight=22.8cm]{geometry}
\usepackage{bm,epstopdf,epsfig,amsmath,amssymb,amsfonts,colordvi,latexsym,
comment,cancel,verbatim,%extarrows,
slashed}
\usepackage{epsfig,bbm}
\usepackage{graphicx,graphics}
\usepackage[font=md,captionskip=8pt]{subfig}
\usepackage[usenames,dvipsnames]{color}
\usepackage[noadjust]{cite}
\usepackage{xcolor}  
\usepackage{setspace}

\newcommand{\qq}{q_{c}}
\newcommand{\rr}{r_{c}}
\newcommand{\h}{\sigma}

\newcommand{\sg}{\sqrt{g}}

\newcommand{\w}{\omega}
\usepackage[utf8]{inputenc}
\allowdisplaybreaks
\newcommand{\cN}{{\cal N}}
\newcommand{\cO}{{\cal O}}   

\newcommand{\cR}{{\tilde R}}

\newcommand{\cT}{{\cal T}}

\newcommand{\ra}{\rightarrow}

\newcommand{\be}{\begin{equation}}
\newcommand{\ee}{\end{equation}}
\newcommand{\bea}{\begin{eqnarray}}
\newcommand{\eea}{\end{eqnarray}}
\newcommand{\Ra}{\Rightarrow}

\newcommand{\baa}{\begin{array}}
\newcommand{\eaa}{\end{array}}

\long\def\symbolfootnote[#1]#2{\begingroup
\def\thefootnote{\fnsymbol{footnote}}\footnote[#1]{#2}\endgroup}
\begin{document} 
\begin{flushright}
%{ \today}
\end{flushright}
\bigskip\medskip
\thispagestyle{empty}
\vspace{2.7cm}
\begin{center}

{\Large \bf  Weyl $R^2$ inflation

\bigskip
with  an  emergent Planck scale}

\vspace{1.cm}

 {\bf D. M. Ghilencea} \symbolfootnote[1]{E-mail: dumitru.ghilencea@cern.ch}

\bigskip
{\small Department of Theoretical Physics, National Institute of Physics
 
and Nuclear Engineering, Bucharest\, 077125, Romania}
\end{center}

\bigskip
\begin{abstract}
\begin{spacing}{1.07}
\noindent
We study inflation in Weyl gravity.
The original Weyl quadratic gravity,   based  on  Weyl conformal geometry, 
is a  theory invariant under the  Weyl symmetry of {\it gauged}  scale transformations. 
In this theory the Planck scale ($M$)  emerges as the scale  where this  symmetry 
 is broken spontaneously  by a geometric Stueckelberg mechanism, to Einstein-Proca  
action for the  Weyl  ``photon'' (of mass  near $M$). With this action as a 
``low energy'' broken phase of  Weyl  gravity,   century-old criticisms of the latter
 (due to non-metricity) are avoided.  In this context,  inflation with 
field values above $M$ is  natural, since this is just a  phase transition 
scale from Weyl gravity (geometry) to Einstein gravity  (Riemannian geometry),
 where the massive Weyl photon decouples.  We show that  inflation in Weyl gravity 
coupled to a scalar field has   results  close to  those in  Starobinsky model 
(recovered for vanishing non-minimal coupling),  with a mildly smaller tensor-to-scalar ratio   ($r$).
Weyl gravity predicts a specific, narrow range  $0.00257 \leq r\leq 0.00303$, for a spectral
index $n_s$ 
within experimental bounds at $68\%$CL and   e-folds number $N\!=\!60$.  
This range of values will soon be reached  by CMB experiments and provides a test 
of Weyl gravity. Unlike in the Starobinsky model, the prediction for $(r, n_s)$ is not 
affected by unknown higher dimensional curvature 
operators (suppressed by some large mass scale) 
since these are forbidden by the Weyl gauge symmetry.
\end{spacing}
 \end{abstract}

\newpage

\section{Motivation}

There is a renewed  interest in studying scale invariant models for
 physics beyond Standard Model (SM)  and cosmology.  This symmetry may also be
present at the quantum level \cite{Englert,SM0,G1,G2}.
All scales (including the scale of ``new physics''  beyond SM)
are generated spontaneously  by field vev's and 
 this symmetry may even preserve  a classical hierarchy of scales
\cite{SM0,G1,G2,G3,G4,M2,SM1,SM2,SM3,Foot}.
In cosmology there are  global \cite{C1,C2,C3,C4,C5,C6,C7,C8,C10,C11,K1,AG1,Karam,Anton,GGR1,GGR2,GGR3}
or local \cite{C12,Oh,TH2,Moffat,Nishino,Smolin,RP1,dg1,dg2,dg3,P,C9} 
scale invariant alternatives to gravity with  spontaneous breaking that
generates the  Planck scale by a non-minimal coupling.

In this paper we study inflation in a theory with {\it gauged}  scale invariance. 
The theory considered is the original   Weyl gravity \cite{Weyl1,Weyl2,Weyl3},
based on Weyl conformal geometry \cite{Scholz}. This   symmetry is
also referred to as Weyl gauge symmetry and its associated  gauge boson is
called hereafter  Weyl ``photon''.
This theory has   {\it no}  fundamental scale    (Planck scale, etc),
 forbidden  by its symmetry. Weyl action is \cite{Weyl1,Weyl2,Weyl3}
\bea\label{WG}
L_0=\sqrt{g}\,\Big\{
\frac{\xi_0}{4!} \,\tilde R^2 -\frac{1}{4 q^2} \,F_{\mu\nu}^2\Big\}
\eea
Each term is Weyl gauge invariant (see later, eq.(\ref{ct}));
 $\tilde R$ is the scalar curvature of {\it Weyl geometry}; $F_{\mu\nu}$
is the field strength of the Weyl ``photon'' $\w_\mu$, with coupling $q$.
In addition to $L_0$, one can also include an independent  Weyl-tensor-squared term of Weyl 
geometry, allowed by the symmetry and considered later.
A topological Gauss-Bonnet term of Weyl geometry may be present too, not relevant here.
However, higher dimensional  operators ($\tilde R^4$, etc) 
cannot be present in $L_0$  since there is no scale to suppress them.

As recently shown in \cite{dg1,dg2} scale dependence in Weyl gravity (\ref{WG}) emerges 
spontaneously after a geometric version of Stueckelberg   breaking mechanism \cite{S}
 of   Weyl gauge symmetry: the dilaton, which is  the Goldstone mode of this 
symmetry (and spin 0 mode propagated by $\tilde R^2$)  is absorbed by the
 Weyl ``photon'' which thus becomes massive.
Denoting by  $R$ the Ricci scalar of {\it Riemannian
 geometry}, action (\ref{WG}) becomes \cite{dg1,dg2}:
\bea\label{EP}
L_0=\sqrt g\,\Big\{-\frac12 M^2\,R
- \frac{3\,M^2}{2\,\xi_0} -\frac{1}{4} F_{\mu\nu}^2 +\frac34 q^2 M^2 \w_\mu\w^\mu\Big\}.
\eea
The Planck scale  $M$ is fixed   by the dilaton vev and the Weyl 
``photon'' acquired a  mass $\sim q M$ near the Planck scale (for $q$ not too small).
Therefore the Einstein-Proca action (\ref{EP}) is just
 a {\it ``low-energy'' broken phase}  of Weyl gravity:
it is obtained   after  ``gauge fixing''  the   Weyl gauge symmetry.  
This involves fixing the dilaton  vev which  in a FRW universe   is  a
 dynamical effect \cite{Fe}. Given its equivalence  to  action (\ref{EP}),  Weyl 
action (\ref{WG})   avoids  long-held criticisms since Einstein \cite{Weyl2} 
 related to non-metricity effects  due to $\w_\mu$, e.g. the changing of atomic lines spacing,
that can be safely ignored since  $\w_\mu$ decouples near Planck scale $M$.

Having Einstein gravity  as a ``low-energy'' limit of Weyl action  motivates
us to study inflation in Weyl gravity.  Moreover, the presence of $\tilde R^2$ 
in (\ref{WG}) points to similarities to successful Starobinsky inflation \cite{Sta}.
And since  Planck scale is  just a phase transition scale, 
field values above $M$ are natural in Weyl gravity. 
This is relevant for inflation where such  values
are common  but harder to accept in models where  Planck scale 
is the physical cutoff. % of the theory. 

\section{Weyl gravity and inflation}

To study Weyl inflation we review briefly the  action of Weyl gravity coupled  to a   
scalar  field $\phi_1$,  see \cite{dg1,dg2} for a detailed analysis\footnote{
Unlike scalars, fermions do not couple to Weyl ``photon'' \cite{Kugo,dg3,Moffat,Nishino}
in the absence of torsion, as here.}.  The  Lagrangian is 
\medskip
\be\label{ll1}
L=\sg\,\Big[\,
\frac{\xi_0}{4!}\,\cR^2 - \frac{1}{4\, q^2}\,%g^{\mu\rho} g^{\nu\sigma}
F_{\mu\nu}^2 % F_{\rho\sigma}
\Big]
- \frac{\sg}{12} \,\xi_1\phi_1^2\,\cR
+
\sg\,\Big[\frac12 \,g^{\mu\nu}\,\tilde D_\mu\phi_1 \tilde D_\nu \phi_1 -
 \frac{\lambda_1}{4!}\phi_1^4\Big],
\ee

\medskip\noindent
with couplings $\xi_0$, $\xi_1$, $\lambda_1$;
$F_{\mu\nu}=\partial_\mu\w_\nu-\partial_\nu\w_\mu$ (no torsion).
$\tilde D_\mu$ is  a Weyl-covariant derivative 
\bea
\label{ST}
\tilde D_\mu \phi_1&=&(\partial_\mu-1/2\,\w_\mu)\,\phi_1,
\\
\cR &=&
R-3 \,   D_\mu \w^\mu -\frac32  \, \w^\mu \w_\mu,
\eea
with the
 scalar curvature of Weyl geometry ($\tilde R$) related to its Riemannian counterpart ($R$)
  as above. 
Lagrangian (\ref{ll1}) is invariant under a Weyl  gauge transformation $\Omega(x)$:
\bea
\label{ct}
 \hat g_{\mu\nu} =  \Omega\, g_{\mu\nu},\qquad
\hat\phi_1  =  \frac{1}{\sqrt \Omega}\,\phi_1, 
\qquad
\hat\w_\mu =\w_\mu-\,\partial_\mu\ln\Omega,
\quad\Ra\quad
\hat \cR=\frac{1}{\Omega}\,\cR.
\eea
where $\omega_\mu$ is the Weyl gauge field.
We also have $\sqrt{\hat g}\!=\! \Omega^2\! \sg$,\,  $g\!\equiv\!\vert\det g_{\mu\nu}\vert$,
$\hat g^{\mu\nu} =  \Omega^{-1}\, g^{\mu\nu}$,  and metric $(+,-,-,-)$. Our conventions are
those of \cite{R}.

Unlike  the Riemannian scalar curvature ($R$),  $\cR$  is 
computed from a Weyl connection {\it invariant} under (\ref{ct}) and
it  transforms covariantly,  due to  the inverse metric  
entering its definition. 
With this observation, one  sees the advantage of Weyl formulation (i.e. using $\cR$) instead of the 
Riemannian language (using $R$) and why $L$  is invariant under (\ref{ct}).

One ``linearises'' (\ref{ll1}) by using an auxiliary field $\phi_0$
to  replace  $\cR^2\!\ra\! -2\phi_0^2 \cR -\phi_0^4$ and to  obtain a classically
equivalent action. Indeed the equation of motion for $\phi_0$ has  solution $\phi_0^2=-\cR$,
 which when used back in the action recovers (\ref{ll1}).  
Given this, $\phi_0$ transforms as any other scalar field. 
Therefore
\medskip
\be\label{tt}
L=\sg\,\Big[
-\frac{1}{4\,q^2} F_{\mu\nu}^2 - \frac{1}{12} (\xi_0\,\phi_0^2+\xi_1\phi_1^2)\,\cR
+ \frac12 \,g^{\mu\nu}\,\tilde D_\mu\phi_1 \tilde D_\nu \phi_1 -
 \frac{1}{4!} (\lambda_1\phi_1^4+\xi_0\phi_0^4)\Big],
\ee

\medskip\noindent
Next, we must fix the gauge  which we do by a particular
 Weyl  ``gauge-fixing''  transformation~(\ref{ct}) of  $\Omega=\rho^2/M^2$
and  $\rho=(1/6) (\xi_1\phi_1^2+\xi_0\phi_0^2)$ with $M$ some scale (see later).
We find
\medskip
\be\label{W3}
L=
\sqrt{\hat g}\, \Big[
-\frac{1}{2} M^2\,\hat{R}+\frac{3}{4} \, M^2\,\hat\w_\mu\hat\w^\mu
-\frac{1}{4 q^2} \hat F_{\mu\nu}^2 
+\frac{\hat g^{\mu\nu}}{2}\hat{\tilde D}_\mu\hat\phi_1 \hat{\tilde D}_\nu \hat\phi_1
-\hat V
\Big],
%\bigskip
\ee

\medskip\noindent
with  $\hat R$ the Riemannian scalar curvature and
% ${\hat{\tilde D}}_\mu\hat\phi_1=(\partial_\mu-q/2\,\hat\w_\mu )\hat\phi_1$ and with 
\medskip
\bea\label{fff}
\hat V= \frac{3 M^4}{2\,\xi_0} 
\Big[1-\frac{\xi_1\hat\phi_1^2}{6 \,M^2}
\Big]^2
+
\frac{\lambda_1}{4!}\,\hat\phi_1^4.
\eea

\medskip
Therefore, as  detailed in \cite{dg1,dg2},
the Weyl ``photon'' has become massive via a Stueckelberg mechanism
by ``absorbing''  the field $\ln\rho\sim \ln\Omega$, eq.(\ref{ct}), 
 of the radial direction in the field space $\phi_0$, $\phi_1$.
This is the Goldstone (dilaton) mode since  under (\ref{ct})  $\ln\rho$ has a
shift symmetry $\ln\rho^2\ra \ln\rho^2 -\ln\Omega$.
Weyl gauge  symmetry is now spontaneously broken
and Einstein gravity is recovered as a ``low-energy'' {\it broken phase} of Weyl gravity
(this is consistent with the fact that  Riemannian-based $R^2$-gravity 
is equivalent to Einstein gravity plus a massless mode \cite{K1}
eaten here by the Weyl ``photon'').
The scale of breaking and the mass of Weyl ``photon'' is $\sim q M$, with $M$ identified 
with the Planck scale. $M$ is itself fixed by the vev of the radial direction  which in a FRW universe
is fixed dynamically \cite{Fe} following a conserved  current which drives the dilaton field
to a constant value.  The Planck scale and the mass of $\w_\mu$ are thus determined by the dilaton vev.
Fields values above  $M$ {\it are naturally allowed} here since $M$ is just
 a phase transition scale in the theory.

The Weyl-covariant derivative acting on $\hat\phi_1$ in (\ref{W3})  is a remnant of the initial
Weyl gauge  symmetry, now broken; one would like to decouple  $\partial_\mu\hat\phi_1$ from $\w^\mu$ 
in order to study inflation;  for a  ``standard''  kinetic term  for $\hat\phi_1$,  a 
field redefinition is used
\medskip
\bea\label{ttprime}
%%\hat\w_\mu^\prime =\hat w_\mu- \frac{1}{q}\partial_\mu \ln \big(\hat \phi_1^2+6 M^2\big),\qquad
\hat\w_\mu^\prime =\hat w_\mu- 
\partial_\mu \ln \cosh^2 \frac{\h}{M\sqrt 6},\qquad
\hat \phi_1= M\sqrt{6}\,\sinh\Big[\frac{\h}{M\sqrt 6}\Big]
\eea
to find 
\be\label{fine}
L=\sqrt{\hat g}\, \Big\{
-\frac{1}{2} M^2\,\hat{R}
+ \frac34  M^2 \cosh^2 \Big[\frac{\h}{M\sqrt 6}\Big] \,\hat\w^\prime_\mu\hat\w^{\prime\,\mu}
-\frac{1}{4 q^2} % \hat{g}^{\mu\rho} \hat{g}^{\nu\sigma} 
\hat F^{\prime\, 2}_{\mu\nu}  % \hat F^\prime_{\rho\sigma}
+\frac{\hat g^{\mu\nu}}{2}\partial_\mu\h \partial_\nu \h
- V \Big\}.
\ee
Finally, one rescales above  $\hat \w_\mu^\prime\ra q \hat \w_\mu^\prime$, 
for a canonical gauge kinetic term.
The potential is
\bea\label{hatV}
V&=&
%\frac{3}{2} \frac{M^4}{\xi_0}
V_0\,\,
 \Big\{\,
\Big[
1-\xi_1\sinh^2 \frac{\h}{M\sqrt 6}\Big]^2
+
\lambda_1\xi_0\,  \sinh^4\frac{\h}{M\sqrt 6}
\Big\},
\qquad
V_0=\frac32\,\frac{M^4}{\xi_0}.\label{ov1}
\eea

\medskip\noindent
$V$  encodes the effect of the initial presence of the  Weyl ``photon''.
The potential has a minimum due to the non-minimal
gravitational coupling $\xi_1>0$ and this is relevant for inflation; we assume
$\h$ as the inflaton, with $V$ its potential.

\section{Results}\label{result}

\begin{figure}[t!]
\begin{center}
\includegraphics[height=0.46\textwidth]{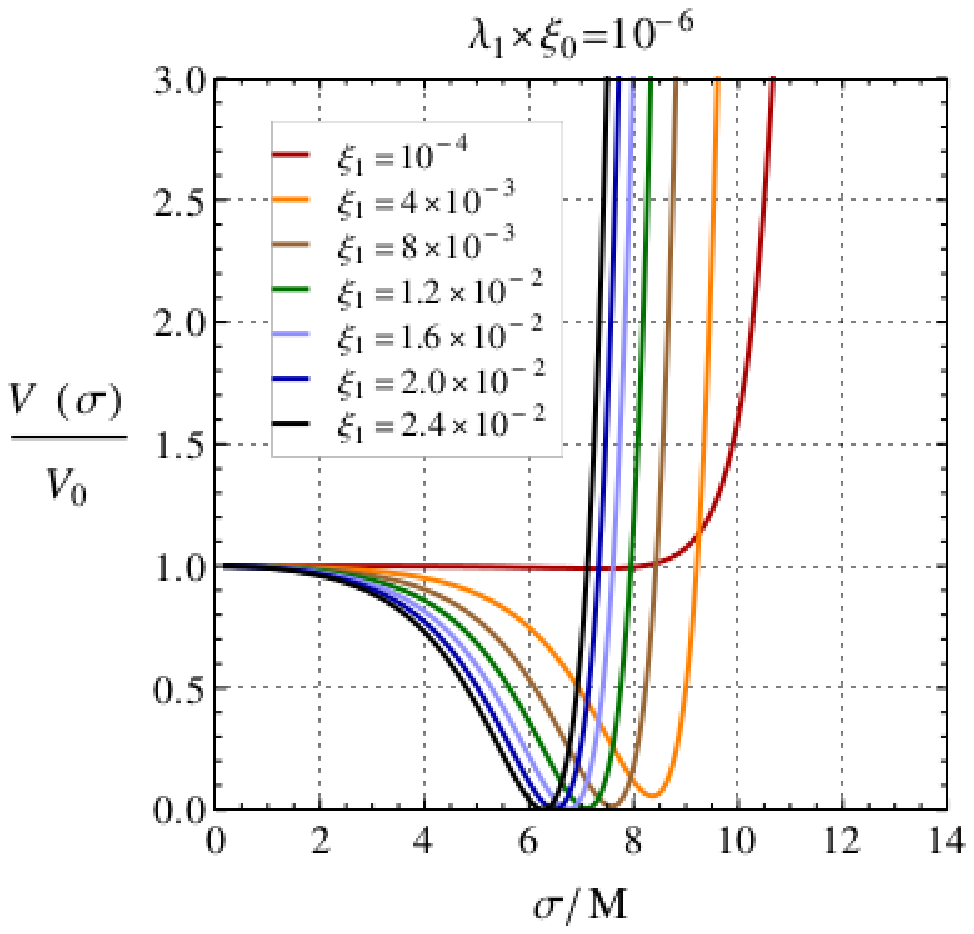} 
\includegraphics[height=0.46\textwidth]{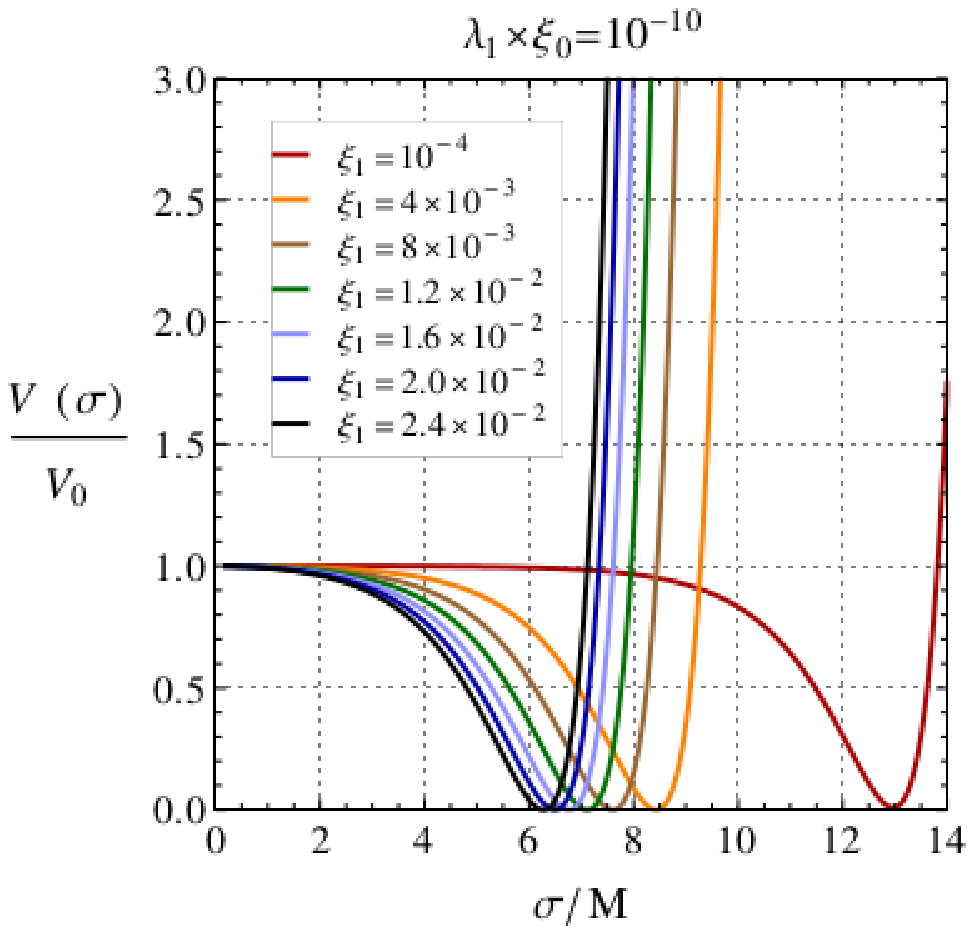}
\end{center}
\caption{The potential $V(\sigma)$ for two fixed values of $\lambda_1\xi_0$ and 
with different   $\xi_1$. 
The flat region is wide for a large range of $\h$, with the width controlled  by
$\gamma\sim 1/\sqrt\xi_1$ while its height is $V_0\propto 1/\xi_0$. 
We have  $V/V_\text{min}\propto\xi_1^2/(\lambda_1\xi_0)$.
}\label{fig1}
%\end{figure}
% \begin{figure}[h!]
  \begin{center}
  \quad
\includegraphics[height=0.43\textwidth]{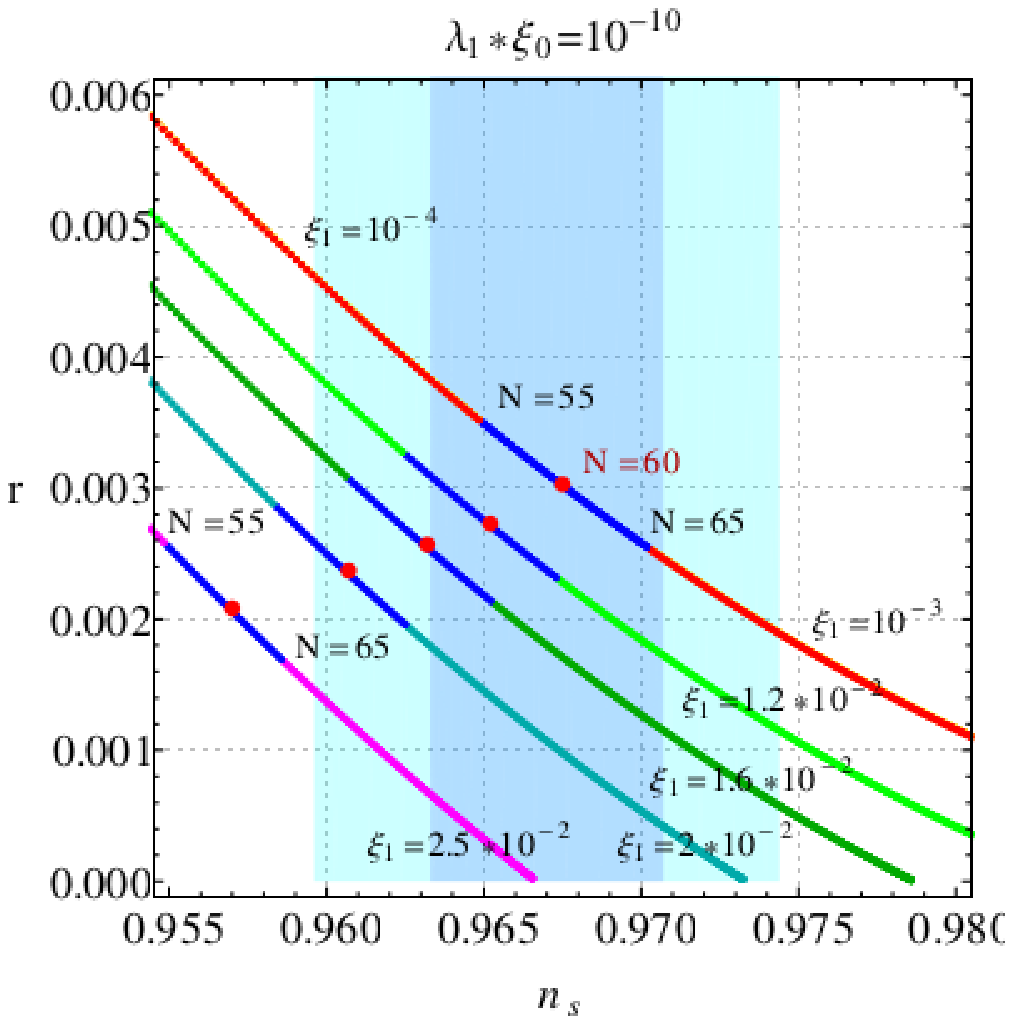}  
\qquad
\includegraphics[height=0.43\textwidth]{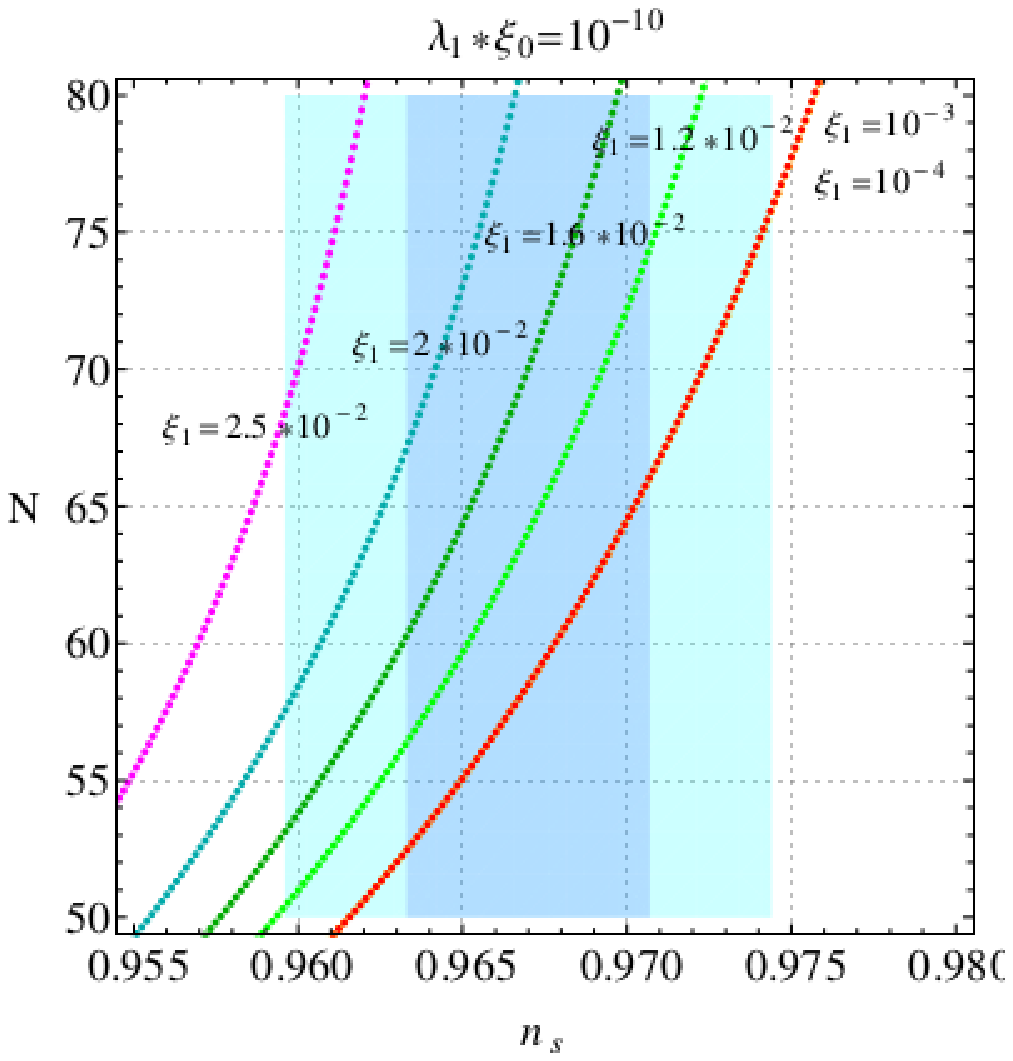} 
 \end{center}
  \caption{ 
The spectral index $n_s$  versus  $r$ (left plot)
and $n_s$ versus  $N$  (right plot),
for models with   $\lambda_1 \xi_0=10^{-10}$ and with
different $\xi_1$. $N$ varies along each curve; in the left plot, on each curve of fixed
 $\xi_1$,  the dark-blue  curve has  $N=55$ and $N=65$ at its left and right ends, 
respectively. The red dots correspond to $N=60$.
 The curves for $\xi_1=10^{-3}$ and $\xi_1=10^{-4}$ are nearly identical (saturated bound as $\xi_1\ra 0$) 
in both plots. Blue (light blue) regions correspond to  $n_s$ given by
$n_s=0.9670 \pm 0.0037$ at 68\,$\%$ CL  (95\,$\%$ CL) 
respectively \cite{planck2018}.
 The constraint  $r\!<\!0.07$  is easily satisfied. The value of $r$ is very small,
below that of  Starobinsky case: for $N=60$ we have $r\leq  0.00303$ (from top curve,
 saturated bound as $\xi_1\!\ra\! 0$) and $r\!>\!0.00257$ from the lower bound on $n_s$ ($68\%$ CL)
with $\xi_1=1.6\times 10^{-2}$.
 }
  \label{fig2}
\end{figure}

$V$ of (\ref{hatV}) is largely controlled by $\xi_1$ and the
combination $\lambda_1\xi_0$.  Its minimum is at:
%\medskip
\bea
\frac{\h_\text{min}}{M\sqrt 6}
=\ln\Big[\gamma+\sqrt{1+\gamma^2}\Big],\quad \gamma=\Big[
\frac{\xi_1}{\xi_1^2+\xi_0\,\lambda_1}\Big]^{1/2},
\qquad V_{\text{min}}=
\frac32\,\frac{M^4\,\lambda_1}{\lambda_1\xi_0 +\xi_1^2}.
\eea
The potential is shown in Figure~\ref{fig1} in function of the field $\h$, for different
{\it perturbative} values of
the non-minimal coupling  $\xi_1$, with two fixed values of the product ($\xi_0\lambda_1$).
Note that:

\noindent
{\bf a)} For $(\lambda_1\xi_0)$ and  $\xi_1$  small enough, $V$ is  constant
 $V\approx V_0\sim 1/\xi_0$ and controlled by $\xi_0$.

\noindent
{\bf b)} Inflation begins in the region $V=V_0$ and lasts a number of e-folds that
 depends on the width of the flat region i.e. on the position of $\h_\text{min}\propto
\gamma$.  If  $\lambda_1\xi_0\ll \xi_1^2$ then $\gamma\sim 1/\sqrt \xi_1$ so reducing
$\xi_1$ will extend the flat region.

\noindent
{\bf c)} From the condition the initial energy be larger than  at the end of inflation,
 $ V_0\gg V_{\text{min}}$ then $\lambda_1\xi_0 \ll \xi_1^2$  and also $V_\text{min}\approx 0$,
 as seen from eq.(\ref{hatV}) and  the second plot in  Figure~\ref{fig1}.

Constraints on the parametric space are found
from the normalization of CMB anisotropy
$V_0/(24\pi^2 M^4\,\varepsilon_*)=\kappa_0, \quad \kappa_0\equiv 2.1\times 10^{-9}$
\cite{planck2018}  where $\epsilon_*$ is the slow roll parameter. 
With tensor-to-scalar ratio  $r=16 \epsilon_*$ and  $r<0.07$ \cite{planck2018}  then 
$\xi_0=1/(\pi^2 \,r\,\kappa_0)\geq  6.89\times 10^{8}$.
In conclusion we have the parametric constraints
\bea\label{xi0}
\lambda_1 \xi_0\ll \xi_1^2,\qquad \xi_0\geq  6.89\times 10^{8}.
\eea
A large $\xi_0$ is always compensated by choosing an ultraweak value of 
$\lambda_1\!\ll \xi_1^2/\xi_0\sim 10^{-9}\xi_1^2$
 so eq.(\ref{xi0}) is respected for a chosen  $\xi_1$ (note the coupling of 
$\tilde R^2$ is $1/\xi_0$ and is in the perturbative regime).
 We shall use these constraints to predict the spectral index $n_s$ and  $r$.

The potential slow-roll parameters are
\medskip
\bea\label{eps}
\qquad\quad
\epsilon
=
\frac{M^2}{2}\,\Big\{\frac{ V^{\prime}}{V}\Big\}^2
=
\frac{1}{3} \frac{\sinh^2(2\tilde \sigma) \big[-\xi_1+
(\lambda_1\xi_0  +\xi_1^2)\,\sinh^2\tilde \sigma\big]^2}{
\big[1 -2\,\xi_1 \sinh^2\tilde\sigma +(\lambda\xi_0 +\xi_1^2)\,
\sinh^4\tilde\sigma\big]^2},
 \quad 
 \tilde\sigma\equiv \frac{\sigma}{M\sqrt 6}, %\quad \tilde\lambda=\lambda_1\,\xi_0.
\eea
%\medskip\noindent
and
%\medskip
\bea\label{eta}
\eta
=
M^2\,\frac{V''}{ V}=
\frac{V_0}{3}
\frac{(\lambda_1\xi_0+\xi_1^2) \,\cosh (4\tilde\sigma)
 -(2\,\xi_1+\lambda_1\xi_0 +\xi_1^2) \,
\cosh (2\tilde\sigma)}{1 -2\xi_1 \sinh^2\tilde\sigma +(\tilde \lambda +\xi_1^2)\,
\sinh^4\tilde\sigma}.
\eea

\medskip\noindent
For $\lambda_1\xi_0\ll \xi_1^2$ and $\xi_1\ll 1$, slow roll conditions are met,
 $\epsilon, \eta\ll 1$, as seen from a numerical analysis. 
Further, the number $N$  of e-folds is
\medskip
\bea\label{NN}
N&=&\frac{1}{M^2}\int_{\sigma_e}^{\sigma_*}
d\sigma \,\frac{V(\sigma)}{V^\prime(\sigma)}=\cN(\sigma_*)-\cN(\sigma_e)
\eea
with
\bea\label{cs}
\cN(\sigma)
= 
c_1\,\ln\cosh\frac{\sigma}{M\sqrt 6}
+
c_2\,\ln\Big[2 \,(\lambda_1\xi_0+\xi_1^2)\,\sinh^2 \frac{\sigma}{M\sqrt 6}-2\,\xi_1\Big]
+c_3\ln\sinh\frac{\sigma}{M\sqrt 6}%\Big\vert_{\sigma_e}^{\sigma_*},
\eea
%\medskip\noindent
and
\medskip
\bea\label{c123}
c_1=\frac{3}{2}\,\frac{\lambda_1\xi_0+(1+\xi_1)^2}{\xi_1+\lambda_1\xi_0+\xi_1^2},
\quad
c_2=\frac{3\,\lambda_1\xi_0}{4\,\xi_1 \,(\xi_1+\lambda_1\xi_0+\xi_1^2)},
\quad
c_3=\frac{3}{-2\,\xi_1}.
\eea

\medskip\noindent
Above $\h=\h_*$ is the value  at the horizon exit. Inflation ends at
 $\h=\h_e$ found from $\epsilon=1$.

Eqs.(\ref{eps}) to (\ref{c123})  are used for a numerical study of the scalar spectral index $n_s$, 
the tensor-to-scalar ratio  $r=16\,\epsilon_*$ and number of e-folds $N$, in terms of 
$\xi_1$ and $\lambda_1\xi_0$. We have
\medskip
\bea\label{ns}
n_s=1+ 2\,\eta_*-6\,\epsilon_*,
\eea
%\medskip\noindent
where the subscript $*$ stands for  $\h=\h_*$.

Before discussing  numerical results, we present  simple analytical results 
for $n_s$ and $r$ in the limit $\lambda_1\xi_0\ll \xi_1^2\ll 1$.
Therefore, we can keep only the leading term in expansion in $(\lambda_1\xi_0)$,
then expand in $\xi_1$. These results depend on $\xi_1$ only in this approximation. 
We find 
\medskip
\bea\label{eqeps}
\epsilon&=& %\Big\{\,
\frac13\, \,\xi_1^2\, \sinh^2 \frac{2\h}{M\sqrt 6} +\cO(\xi_1^3)\,
%\Big\}+\cO(\lambda_1\xi_0)
\\
\eta&=&%\Big\{
-\frac23\,\xi_1 \,\cosh\frac{2\h}{M\sqrt 6}+\cO(\xi_1^2)\,
% \Big\}+\cO(\lambda_1\xi_0)
\label{eqeta}
\eea

\medskip\noindent
valid up to  $\cO(\lambda_1\xi_0)$ corrections.
Therefore
%\medskip
\bea\label{nsns}
\qquad
n_s\, =\, 
% 1 +2 \eta_* -6 \epsilon_*=
%\Big\{\,
1 -\frac43\,\xi_1 \,\cosh\frac{2\h_*}{M\sqrt 6}+\cO(\xi_1^2)\,
%\Big\}+\cO(\lambda_1\xi_0).
\eea

\medskip
The value of $n_s$ is controlled by $\eta$ in leading order  $\cO(\xi_1)$ while $\epsilon$ 
contribution is subleading $\cO(\xi_1^2)$; hence we have  a small tensor-to-scalar ratio
 $r=16\,\epsilon_*$.  We then have
\bea\label{rns}
r= 3\, (1-n_s)^2 +\cO(\xi_1^2)
\eea
This is an approximate result (valid only for smallest $\xi_1$) 
for the top curve shown in the left plot in Figure~\ref{fig2} (this figure actually uses the exact 
numerical results discussed later). As an example, if  $n_s=0.968$  then $r\approx 0.003$.
Increasing $\xi_1$ should reduce $n_s$ of (\ref{nsns}) but
 there is implicit $\xi_1$-dependence in  $\h_*$, computed below. First, we have
\medskip
\bea\label{Ne}
\cN(\sigma)=
%\Big\{
-\frac{3}{4\,\xi_1}\ln\tanh^2\frac{\h}{M\sqrt 6}
+\frac{3}{4} \ln\cosh^2\frac{\h}{M\sqrt 6}
+\cO(\xi_1^2)
% \Big\}+\cO(\lambda_1\xi_0).
\eea

\medskip\noindent
Inflation ends at  $\h=\h_e$ where
\medskip
\bea\label{o1}
\sinh^2\frac{2\h_e}{M\sqrt 6}=\frac{3}{\xi_1^2}+\cO(\xi_1).
\eea

\medskip\noindent
For example: $(\h_e, \xi_1)=(12.8 M, 10^{-4})$; 
% ($9.98 M, 10^{-3}$); $(8.28 M, 4\!\times\!10^{-3})$,
$(7.16M,10^{-2})$.  Then from (\ref{Ne}), (\ref{o1})
\medskip
\bea\label{o2}
\cN(\h_e)=\Big[\frac{\sqrt{3}}{2}+
\frac38 \ln\frac34\Big]
-\frac34 \ln\xi_1+\frac{\sqrt 3}{4}\xi_1
+\cO(\xi_1^2).
\eea

\medskip\noindent
Using this result  we find an iterative solution $\h_*$   to eq.(\ref{NN}) 
for a fixed $N$:
%\medskip
\bea\label{tan}
\tanh^2\frac{\h_*}{M\sqrt 6}&\approx & 1-
\frac{4\xi_1}{3}\,\overline N,
\eea
where
\bea
\overline N&=&\Big\{ N+\cN(\h_e)+\frac34 \ln\Big[ \frac{4\xi_1}{3} \big(N+\cN(\h_e)\big)\Big]\Big\}
+\cO(\xi_1^2).
\eea

\medskip\noindent 
For $N=60$ then   $\overline N \approx 64.1$ for $\xi_1$  between $10^{-4}$
and $10^{-3}$; for $N=55$ then $\overline N\approx 59.06$ for the same $\xi_1$ range.
Finally, using  $\h_*$ of (\ref{tan}) and (\ref{eqeps}), (\ref{nsns}), we have  approximate relations
for $n_s$ and $r=16\,\epsilon_*$:
%\medskip
\bea\label{nnrr}
n_s&=& 1-\frac{2}{\overline N}+ \cO(\xi_1) %\frac{4\,\xi_1}{3}+\cO(\xi_1^2)
\\
r&=& \frac{12}{{\overline N}^2} + \frac{16}{\overline N}\, \cO(\xi_1) 
% -\frac{16\xi_1}{\overline N}+\cO(\xi_1^2)
\label{rrnn}
\eea
%\medskip\noindent
which are consistent with (\ref{rns}) and accurate for $\xi\sim 10^{-3}$ 
for the  $r$ values considered here.

Let us now analyse the  exact numerical values of $n_s$,  $r$, $N$, 
eqs.(\ref{eps}) to (\ref{ns}) and compare them with the experimental data.
Figure~\ref{fig2} summarises the main results of the paper: 
 we presented $n_s$ versus $r$ (left plot) and
$n_s$ versus  $N$ (right plot), for models with  parametric constraint  $\lambda_1 \xi_0=10^{-10}$
 and curves of  different $\xi_1$.
The curves for $\xi_1=10^{-3}$ and $\xi_1=10^{-4}$ or smaller are nearly identical (saturated bound) 
in both plots.  $N$ varies along each curve in the plane $(n_s,r)$, 
and  the range of values from $N=55$ to $N=60$ is shown in dark blue for each curve.
 Regions in blue (light blue) show the experimental values of $n_s$ at $68\%$ ($95\%$ CL), respectively,
where  $n_s=0.9670\pm 0.0037$  ($68\%$ CL) from Planck  2018 (TT, TE, EE + low E 
+ lensing + BK14 + BAO) \cite{planck2018}. These bounds on $n_s$ and $r$ are comfortably 
respected at $68\%$ CL for the values of $\xi_1$ shown.

Let us compare Figure~\ref{fig2} to a result of 
Starobinsky model in which for $N=55$ one has $n_s\approx 0.965$ and $r\approx 0.0034$
\cite{PDG}. In our model, for the same $N$ we also have  $r\approx 0.00345$ 
which is the largest $r$ in the model for $N=55$ (top curve in Figure~\ref{fig2}). 
Therefore the values for $r$ of the  Starobinsky model (recovered for $\xi_1\ra 0$),
are at the upper limit  of those in our model.
This is also indicated by approximate results (\ref{nnrr}), (\ref{rrnn}) which also apply to
Starobinsky model but with replacement  $\overline N \ra N$. With $\overline N>N$ we see that  
for the same $N$, we expect a mildly larger $n_s$ and smaller $r$ than 
in Starobinsky model.

Figure~\ref{fig2} for  $N=60$ gives a lower  bound for 
$r$ from the experimental (lower) value of $n_s$ (at $68\%$ CL) quoted above and 
with $\xi_1=1.6\times 10^{-2}$; 
there is also an upper bound on $r$  from the smallest $\xi_1$ (saturated limit, top red curve)
also giving the largest $n_s$:
\medskip
\bea
N=60,\qquad  0.00257 \leq r\leq 0.00303.
\eea

\medskip\noindent
Similar bounds can be extracted for different $N$.
To conclude,  a small tensor-to-scalar ratio  is predicted by this model. 
Such value will soon be tested experimentally  \cite{CMB,Errard:2015cxa,Suzuki}.

Our predictions used a hierarchy of couplings  $\lambda_1\!\ll\! \xi_1\!\ll\! \xi_0$.
This hierarchy is stable under matter (scalar) quantum corrections to (\ref{ll1})
due to ultraweak  $\lambda_1$ required to satisfy (\ref{xi0}): 
we have  one-loop corrections $\delta \lambda_1\!\propto \!\lambda_1^2/\kappa;$
$\delta \xi_1\!\propto\! (\xi_1+1/6)\lambda_1/\kappa$ and $\delta\xi_0\!\propto\! (\xi_1+1/6)^2/\kappa$,
with $\kappa=(4\pi)^2$, see e.g.\cite{Ghilencea:2018rqg} (Section 3.1).  
Therefore the relative $\lambda_1$ suppression factor can maintain this hierarchy since then
$\vert\delta\lambda_1\vert\ll\vert\delta\xi_1\vert\ll\vert\delta\xi_0\vert$. 
Therefore matter quantum corrections to  $r$ and $n_s$ are small  (for the Starobinsky-Higgs 
model these are well below $2.5\%$ for $r$ and less than $1\sigma$ for $n_s$
for minimal values of $\lambda_1$ used here   \cite{Ghilencea:2018rqg}).

\section{Weyl-tensor corrections to  Weyl  inflation}

The analysis  so far  was based on  $L_0$ of (\ref{WG}) coupled to the inflaton.
A most general  Weyl quadratic gravity action
can contain  an  additional  term allowed by symmetry (\ref{ct}),  \cite{dg1,dg2}
\medskip\bea\label{L1}
L_1=\frac{1}{\eta} \,\tilde C_{\mu\nu\rho\sigma} \tilde C^{\mu\nu\rho\sigma}
=\frac{1}{\eta} \,\Big[ C_{\mu\nu\rho\sigma} \, C^{\mu\nu\rho\sigma} +\frac32\,F_{\mu\nu}^2\Big],
\eea

\medskip\noindent
where $\tilde C_{\mu\nu\rho\sigma}$ ($C_{\mu\nu\rho\sigma}$) denotes the Weyl tensor of 
Weyl (Riemannian) geometry, respectively;  in the second step we used an identity
to rewrite the action in the Riemannian picture.

If present,  $L_1$  modifies  our previous results for $r$.
The analysis proceeds as before, since the
transformations applied to obtain eqs.(\ref{fine}), (\ref{hatV})
 from eq.(\ref{ll1})  leave $L_1$  invariant.  
Therefore, the new Lagrangian is that of eq.(\ref{fine}) plus  $L_1$.
For this new Lagrangian, the corrections to $r$ induced by the Weyl tensor 
were studied in \cite{Baumann}  to which we refer 
the reader for details. The Weyl tensor brings a quadratic action for 
 tensor fluctuations  and a modified sound speed $c_t$ given by
%\bea
$1/c_t^2-1=  (2 H^2)/(\eta M^2)$ \cite{Baumann} (see eqs.(2.10), (2.22), (2.24)).
%\eea
With a slow roll relation $H^2\!\approx V/(3 M^2)$,  $V\!\approx\! V_0$,  
the corrected tensor-to-scalar ratio $\rr$ is then
\bea\label{ttp}
\rr=r\,\Big[ 1 + \frac{8}{\eta\,\xi_0}\Big]^{1/2},
\eea
where $r$ is the value in the absence of $L_1$ while  $n_s$ is unchanged.
The Weyl boson mass is as in eq.(\ref{W3})
but with $q^2\ra \qq^2=q^2/(1-6 q^2/\eta)>0$ where $\qq$ is the new corrected
 coupling of the
 gauge kinetic term which  includes that from  $L_1$; also
$\qq^2\!<\!1$ if $\eta\!<\!0$ or $\eta\!>\!6 q^2/(1-q^2)$.

As expected, eq.(\ref{ttp}) shows that  the  change of $r$ due to $L_1$
 depends on the relative size of the
perturbative couplings of the two quadratic terms,
 $\eta$ versus $1/\xi_0$, and equals $\theta=(\rr-r)/r\approx 4/(\eta\,\xi_0)$
for $\vert \eta\vert \,\xi_0\gg 1$.
Recalling that  $\xi_0> 6.89 \times 10^8$,  a
relative change by $\theta$ requires  $\eta\approx 6\times 10^{-9}/\theta$.
A value $\rr\! <\!r$ ($\rr\! >\!r$)
 corresponds to a negative (positive) $\eta$, respectively.
In the limit $L_1$ is absent (formally $\vert\eta\vert\ra \infty$, $q$ fixed) then
$\rr=r$ and $\qq=q$ which recovers our previous results.
In the limit the gauge kinetic term comes solely from (\ref{L1}) i.e. 
$L_0$ contains only the $\tilde R^2$ term
(formally $q\ra\infty$, $\eta$ fixed) then one has $\eta=-6\qq^2<0$. 

A significant change $\theta$  (e.g. $\theta=50\%$) of an already very small  (and 
well below current experimental bounds) 
tensor-to-scalar ratio needs an ultraweak  $\vert \eta\vert$  (correspondingly
$\eta\!=\!1.2\!\times\! 10^{-8}$),
which induces an instability in the theory below Planck scale. This is because there is a spin-two ghost 
(or tachyonic) state  in $L_1$ \cite{AG1} of  (mass)$^2\propto\eta\, M^2$. 
Avoiding this instability below  Planck scale 
means corrections to  $r$ from $L_1$ are negligible.

\section{Further remarks}

A  potential similar to that in Section~\ref{result} was encountered in a 
previous model \cite{P} with Weyl local symmetry. What are the differences? In our model
 there is no torsion (Weyl connection coefficients are symmetric \cite{dg1,dg2}) but we 
have Weyl gauge symmetry  and non-metricity, while
in the model of \cite{P} only torsion is present.
Further, in \cite{P}  the ``gauge'' field (denoted $\cT_\mu$)  emerges from the  trace
 over  the torsion and replaces our $\omega_\mu$. However, an ansatz is made  
\bea\label{torsion}
\cT_\mu=\partial_\mu\phi
\eea
%\medskip\noindent
with $\phi$ a scalar field.
Under this assumption the model is  Weyl integrable i.e. Riemannian (see e.g.\cite{Q2})
and then non-metricity is absent ($\cT_\mu$ being a ``pure'' gauge field).  Due to (\ref{torsion}) the
gauge kinetic term of $\cT_\mu$ is  vanishing (no dynamics) and can be integrated out.
Therefore,  no geometrical  Stueckelberg mass mechanism can take place.
For this reason a flat (Goldstone) direction remains present in \cite{P} and it
has kinetic mixing with the inflaton. The results then depend on the dynamics of the flat
direction.  This has implications for inflation discussed in \cite{P}, 
where it is shown that a distinct field space geometry
changes the slow-roll plateau,  which can affect inflation.
If the kinetic energy of the Goldstone is large it can dominate and 
a ``kination'' period  predates the slow-roll inflation; this  may have 
additional  consequences  (observable effects in the CMB on large
 angular scales) \cite{P}.
In our case there is no Goldstone (flat direction) left since it was eaten by  the Weyl ``photon'' 
which becomes massive via Stueckelberg mechanism and eventually decouples,
 yet it impacts on the potential (compare (\ref{hatV}) to (\ref{fff}))\footnote{Non-metricity 
effects from $\w_\mu$ are  suppressed  by its large mass $q M$, for $q$ perturbative,
not too small. Current  non-metricity bounds   \cite{Olmo,Lobo} are as low as TeV, but
depend on the model details. The fermions in our model  do not couple to  Weyl ``photon'' 
\cite{dg3,Moffat,Nishino} and  may evade these constraints even for ultraweak $q$.}.

Weyl inflation  has an advantage compared to  the Starobinsky model in that 
it cannot contain higher dimensional/curvature operators like $\tilde R^4/M^4$, etc
of {\it unknown} coefficients that 
could affect significantly the  numerical predictions or the convergence
 of such an expansion (in powers of $\tilde R$), see \cite{edholm} for a discussion.
Unlike in the Starobinsky model,  such higher dimensional
 operators and their corrections are forbidden by the Weyl gauge symmetry. One could think
of such operators being suppressed instead  by (powers of) the dilaton field\footnote{
Such situation is possible 
in quantum scale invariant models in flat spacetime   when scale-invariant
higher dimensional operators are generated at quantum level,
suppressed by powers of the dilaton \cite{G1,G2,G3}.}
(to preserve this symmetry) but this is not
possible since  this field is  already ``eaten'' by the Weyl massive ``photon'',
to all orders in perturbation theory.
Further, given the Weyl gauge symmetry, the  model is allowed by black-hole physics, 
which is not the case of similar models of inflation with only {\it global} scale symmetry 
(global charges can  be eaten by black holes which subsequently evaporate, e.g. \cite{Kallosh}).
Finally, compared to  models of inflation with local scale invariance (no gauging)
that have a   ghost  present when generating Planck scale spontaneously by the
 dilaton vev\footnote{See for example  \cite{Oh,dg3} for a discussion and references.}, 
this problem is not present in Weyl gravity action eq.(\ref{ll1}).

\section{Conclusions}

We examined if  the original Weyl quadratic gravity is suitable for  inflation.
This theory is based on Weyl conformal geometry and its gauged (local) scale
symmetry (also called Weyl gauge symmetry) {\it forbids the presence} of any fundamental
mass scale in the action. 
Its action undergoes spontaneous breaking via geometric Stueckelberg mechanism.
In this way the Weyl ``photon'' of gauged dilatations becomes massive (mass $\sim q M$), 
after absorbing the  Goldstone mode (compensator/dilaton) $\ln\phi_0$ 
which is the  spin-0 mode  propagated by the $\tilde R^2$ term in the action.
The result in the broken phase is the Einstein-Proca action for the 
Weyl ``photon'' and a positive cosmological constant. 
If  the initial action also has a non-minimal coupling $\xi_1$ to an additional scalar ($\phi_1$), 
a scalar potential is found after the Stueckelberg mechanism.  This potential has 
a minimum for non-vanishing scalar vev that is triggered by the gravitational effects 
(non-minimal coupling) and is suitable for  
inflation. Since the Planck scale is {\it emergent} as the scale where Weyl gauge
symmetry is broken, the presence  of field values above this scale, needed for inflation,
is actually natural in Weyl gravity. Moreover, the existence of a non-zero 
vev of the dilaton  (fixing $M$) is actually a dynamical effect  in  a FRW universe. 
The study is also motivated by the fact that
 the action involves the square of the (Weyl) scalar curvature,
which  points to similarities to the successful Starobinsky model.

Our analysis shows that Weyl inflation predicts a specific, small tensor-to-scalar ratio 
($r$) within a narrow range $0.00257 \leq r\leq 0.00303$ for $N=60$ and with $n_s$ within $68\%$ (CL) of 
the experimental value. This range of values  for $r$  will soon be 
tested experimentally; they  are mildly smaller than those for same $N$ 
in the Starobinsky model $M^2 R+R^2$  recovered in the limit of vanishing non-minimal coupling.
Such value for $r$ is also an indirect test of the presence of the Weyl gauge symmetry.

Compared to the Starobinsky model, the Weyl model of inflation has the  advantage 
that it does not contain  higher order curvature terms (e.g. effective operators $R^4/M^4$, etc)
that  modify the predictions or question the convergence of  a series expansion in curvature; 
such operators
 are forbidden in Weyl inflation  by the  underlying Weyl gauge symmetry. This is because this symmetry 
does not allow a mass scale be present in the Weyl action to suppress such  operators, while
the dilaton field that could in principle suppress them (while preserving this symmetry)
was already ``eaten'' by the massive Weyl photon. 
 Another advantage is that the Weyl gauge symmetry of this model 
is also allowed by black-hole physics, unlike the  models with a {\it global}
 scale symmetry, while local scale invariant models (no gauging) have a notorious 
ghost dilaton present,  when generating the Planck scale spontaneously (by the dilaton vev).
Finally, the above predictions  for $r$ and the spectral index $n_s$  are found 
 for  values of the  non-minimal coupling $\xi_1$  in the  perturbative regime.
In this respect the situation  is very different from Higgs inflation 
where a  large coupling $\xi_1$ is actually required.

\vspace{2.8cm}
\noindent
{\bf Note added in proof:} While this work was being typewritten, preprint arXiv:1906.03415 
appeared (10 June) which analyses (section 4) inflation in  this model  by using the two-field 
basis  eq.(\ref{tt}) instead of our one-field  formulation, eq.(\ref{hatV}).  
The results are consistent and complementary.
 
%\newpage
\end{document}